\documentclass[a4paper]{article}
\usepackage{INTERSPEECH2022}
\usepackage{multirow}
\usepackage{pifont}

\title{STCON System for the CHiME-8 Challenge}
\name{
Anton Mitrofanov$^{1,2}$, 
Tatiana Prisyach$^{1}$, 
Tatiana Timofeeva$^{1,2}$,
Sergei Novoselov$^{1,2}$,
Maxim Korenevsky$^{1}$,
Yuri Khokhlov$^{1}$,  
Artem Akulov$^{1}$,  
Alexander Anikin$^{1}$,
Roman Khalili$^{1}$,
Iurii Lezhenin$^{1,3}$,
Aleksandr Melnikov$^{1}$,
Dmitriy Miroshnichenko$^{1}$,
Nikita Mamaev$^{1}$,
Ilya Odegov$^{1}$,
Olga Rudnitskaya$^{1}$,
Aleksei Romanenko$^{1,2}$
}
\address{
  $^1$STCON LLC., Kingdom of Saudi Arabia;
  $^2$ITMO University, Russia;
  $^3$Peter the Great St.Petersburg Polytechnic University, Russia;
}
\email{
\{
mitrofanov-aa,
prisyach,
timofeeva,
novoselov,
korenevsky,
khokhlov,
akulov,
anikin,
khalili,
lezhenin,
melnikov-a,
miroshnichenko,
mamaev-n,
odegov,
rudnitskaya,
romanenko
\}@speechpro.com}

\begin{document}
\maketitle
\begin{abstract}
This paper describes the STCON system for the CHiME-8 Challenge Task~1 (DASR) aimed at distant automatic speech transcription and diarization with multiple recording devices. Our main attention was paid to carefully trained and tuned diarization pipeline and speaker counting. This allowed to significantly reduce diarization error rate (DER) and obtain more reliable segments for speech separation and recognition. 
To improve source separation, we designed a Guided Target speaker Extraction (G-TSE) model and used it in conjunction with the traditional Guided Source Separation (GSS) method.  
To train various parts of our pipeline, we investigated several data augmentation and generation techniques, which helped us to improve the overall system quality.

\end{abstract}
\noindent\textbf{Index Terms}: speech recognition, speaker diarization, WPE, GSS, G-TSE, speaker counting, NSD-MS2S, RIR generator, WavLM, ZipFormer, CHiME-8.

\section{System description }
\begin{figure}[!ht]
  \vspace{-10mm}
\centering
\includegraphics[width=.65\linewidth]{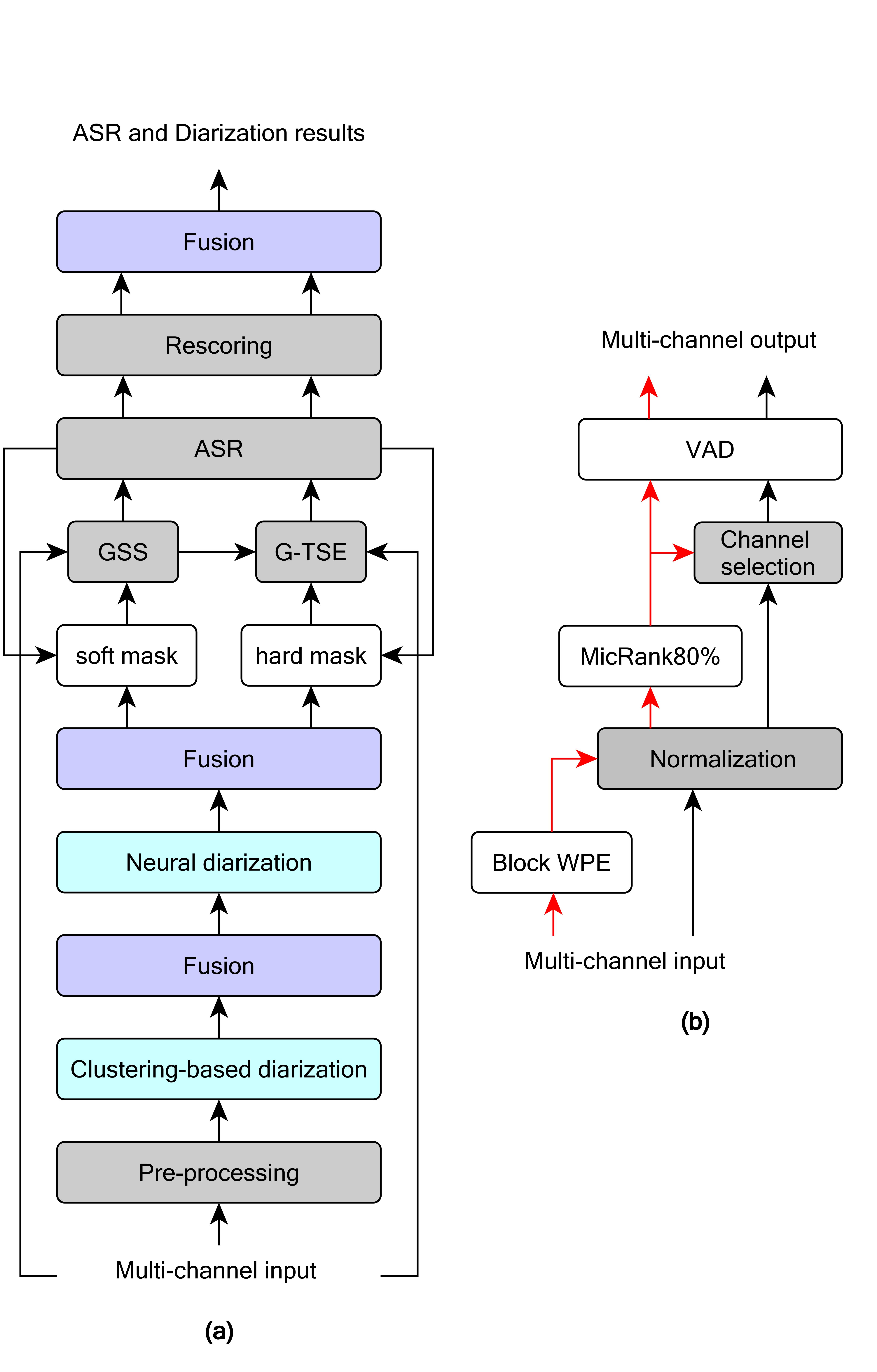}
  \caption{(a) STCON system. (b) Pre-processing diagram.}
  \label{fig:System}
  \vspace{-6mm}
\end{figure}

In general, our system follows a pattern that has worked well in previous CHiME challenges \cite{CHiME6}, \cite{CHiME7}, namely, speaker diarization, source separation and speech recognition. For the current challenge we have paid much attention to the improvement of the diarization and speaker counting. We applied sophisticated multi-step procedure to obtain high-quality clustering-based diarization which is followed by advanced neural diarization system providing accurate speakers' activity bounds. Several variants of Guided Source Separation (GSS) \cite{GSS} as well as Target Speaker Extraction (TSE) \cite{TSE} are then used to extract each speaker's utterances and recognize them with a carefully tuned Automatic Speech Recognition (ASR) models. Finally, ASR results are re-scored with a large Language Model and fused to provide a final speaker-attributed transcription results.
The main components of the system are shown on Fig.~\ref{fig:System}(a).

\section{Speaker diarization}
Since the boundaries of the sentences in the training data are not always accurate and there are pauses inside them, we made a forced alignment of all data based on Individual Headset Microphone (IHM) recordings. This provided a more accurate references for DER measurements, which is important for training good diarization models. All DER results below correspond to these aligned segments, not original ones.

\subsection{Clustering-based diarization}
Basic clustering-based diarization begins with a short but important step of normalization. Each session recording is first clipped by a certain threshold value and then re-normalized to the maximum amplitude. The threshold value is chosen as a predefined percentile of absolute signal values. This allows to get rid of loud knocks and claps. Normalization is applied to both original recordings and ones processed with 2 minute-block Weighted Prediction Error (WPE)  dereverberation \cite{WPE}. Normalized recordings are then processed by an Envelope Variance \cite{MicRank} based channel selection algorithm (hereinafter referred to as MicRank) to select 80\% best channels for each session. MicRank is applied to the WPE-processed recordings and corresponding channels are also selected from the normalized original recordings. All subsequent processing is performed channel-wise.
%
\begin{table*}[!t]
 \caption{Clustering-based diarization results. }
  \label{table:diarization-results}
\vspace{-1mm}
\centering
\begin{tabular}{|l|c|c|c|c|c|c|c|c|c|}
\hline
\multirow{3}{*}{System} & \multirow{3}{*}{max\_spk} & \multicolumn{7}{|c|}{DER / speaker count accuracy} & \multirow{3}{*}{AVG}  \\
\cline{3-9}
& &\multicolumn{2}{|c|}{chime6} & \multicolumn{2}{|c|}{dipco} & \multicolumn{2}{|c|}{mixer6} & \multicolumn{1}{|c|}{notsofar1} & \\ \cline{3-9}
&&dev & eval & dev & eval & dev & eval & dev & \\ \hline
\multirow{2}{*}{baseline [14]} & 4 & 26.8 & - & 24.78 & - & 16.53 & - & - & - \\
 & 8 & 36 & - & 26 & - & 24 & - & - & - \\ \hline
single\_orig$^*$ & 8 & 25.3/0.87 & 31.9/0.87 & 23.7/1 & 20.0/0.85 & 16.3/0.91 & 7.6/1 & 20.0/0.86 & 20.6/0.88 \\ 
single\_wpe$^*$ & 8 & 24.1/1 & 30.9/1 & 22.4/1 & 17.1/0.85 & 12.8/0.97 & 7.7/0.97 & 20.8/0.85 & 19.4/0.87 \\ \hline \hline
fusion & 8 & 23.5/1 & 29.6/1 & 21.4/1 & 17.3/0.85 & 13.0/0.98 & 7.5/1 & 13.0/0.89 & 17.9/0.90 \\ \hline
\multicolumn{9}{l}{$^*$
The best of 6 systems with different parameters $thr$ and VAD segments.}
\end{tabular}
\vspace{-5mm}
\end{table*}

The pre-processing steps are depicted at Fig.\ref{fig:System}(b).

\begin{figure}[!ht]
\centering
\includegraphics[width=.85\linewidth]{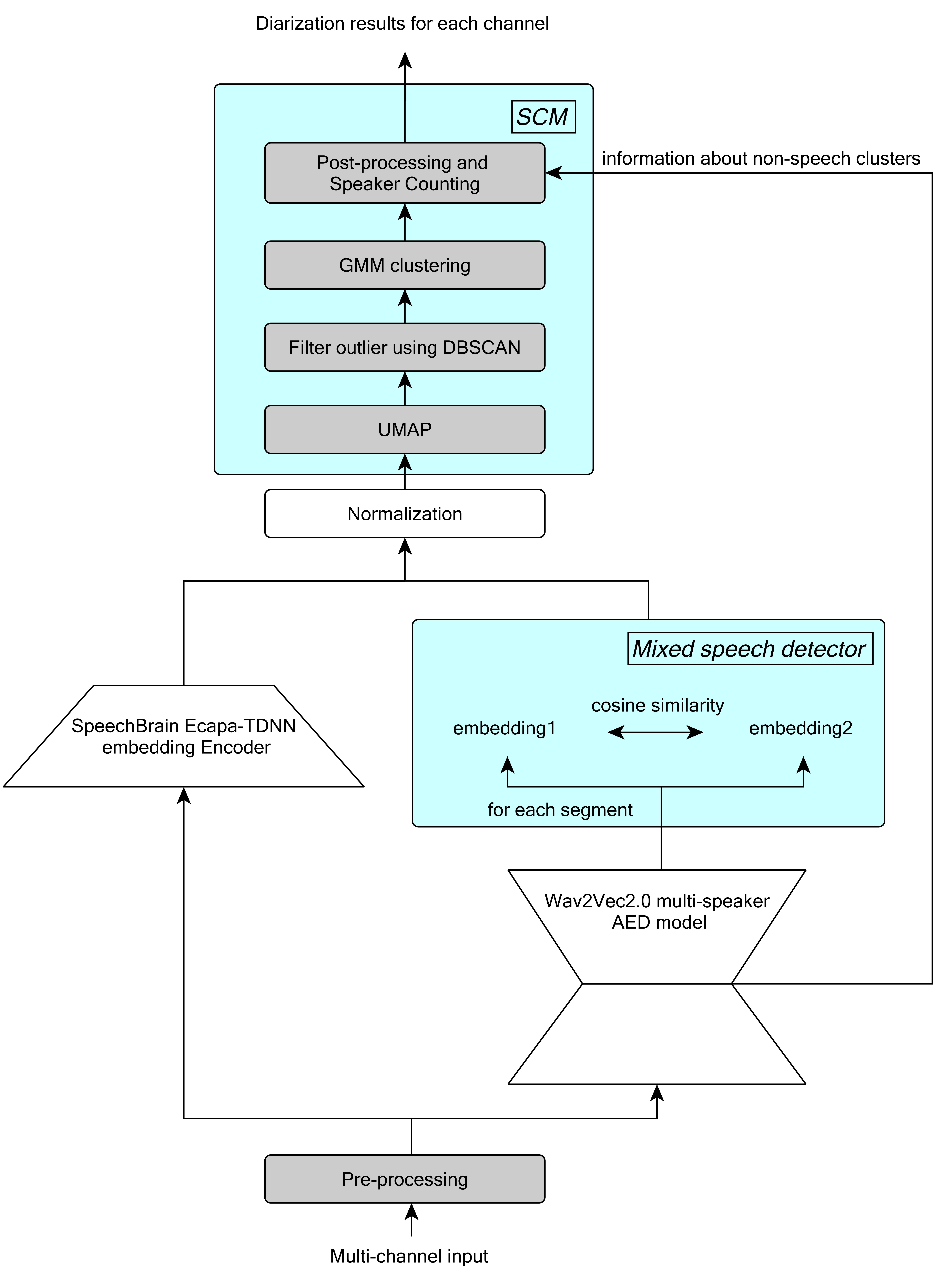}
  \caption{Clustering-based diarization. }
  \label{fig:Clustering-based-diarization}
    \vspace{-5mm}
\end{figure}

Voice Activity Detector (VAD) is applied to select segments where at least one speaker is active. These segments are passed into the speaker counting module (SCM). SCM includes speaker embeddings extraction, clustering and clusters post-processing. Two types of embeddings are used. 
First type is multi-speaker embeddings extracted with an in-house model based on Wav2Vec2.0 XLS-R 53\footnote{https://huggingface.co/facebook/wav2vec2-large-xlsr-53}. 
The model is similar to one described in \cite{w2v2_extractor} and was finetuned on VoxCeleb1,2 data. This model has Attention-based Encoder-Decoder architecture and is trained to extract several embeddings for segments with mixed speech, i.e. to act as a mixed speech detector. Several embeddings extracted from each segment are compared to each other using cosine similarity. If it exceeds a certain threshold, corresponding speakers are merged together.  This processing makes it possible to select a subset of single-speaker frames only. Embeddings of the second type are extracted using the SpeechBrain Ecapa-TDNN model \cite{ecapa}. Both types of embeddings are concatenated and the resulting embeddings are normalized to unit length. UMAP \cite{UMAP} dimensionality reduction downto 12 is applied followed by the GMM clustering \cite{GMMclustering} of embeddings on the pre-selected single-speaker frames. The maximal number of clusters is set to 8 but some of them can be subsequently merged together based on the cosine similarity and rejected based on cluster size threshold $thr$\footnote{Clusters with a size $N$ smaller than $N_{max}/thr$ are rejected}. Besides, a Wav2Vec2.0-based extractor can work as non-speech classifier and remove a cluster if it corresponds to non-speech.
After the number of clusters is determined and fixed, all mixed speech frames are processed. Each such frame is greedily attracted to the cluster whose centroid is the nearest to the corresponding embedding. The scheme of the clustering-based diarization is depicted on Fig.~\ref{fig:Clustering-based-diarization}.

\subsection{Neural diarization (ND)}
\label{subsec-ND}
This module is based on the open-sourced CHiME7 winner solution Neural Speaker Diarizarion using Memory-Aware and Multi-Speaker embeddings (NSD-MS2S)\footnote{https://github.com/liyunlongaaa/NSD-MS2S}. Several models for NSD-MS2S were trained. Each of them was pretrained on a large set of simulated CHiME8-like data and then fine-tuned on all training data from all Multiple Distant Microphone (MDM).

To prepare the simulated dataset, a set of artificial Room Impulse Responses (RIRs) was generated, which are similar to those of CHiME8 data. Five different RIR-classifiers \cite{RIRclassifier} were trained, each on different 100K artificially generated\footnote{https://pypi.org/project/rir-generator} RIRs. Classifiers were trained on Librispeech \cite{Librispeech} data augmented by noises extracted from CHiME8 data. 
All recordings from the CHiME8 training data were classified\footnote{Each RIR-classifier takes a recording and returns a probability distribution of its RIRs in this recording. These distributions were averaged over all oracle segments and all MDM channels for each dataset. The more probable RIR is the more relevant it is for the dataset.} by each of classifier and 20\% RIRs with the highest probability on each subset of CHiME8 data were selected, resulting in 20K selected RIRs per classifier or 100K RIRs in total. The room parameters of the selected RIRs were used to generate a set of multichannel RIRs with 20 sources and 10 receivers per each room. Using these multichannel RIRs, LibriSpeech data and noises extracted from CHiME8 data, the 10-channel diarization dataset of about 2500 hours was prepared with the help of mixture simulation script from BUT\_EEND repository \cite{BUT_EEND}. The statistics of speaker overlaps for the simulation was collected from NOTSOFAR-1.

\begin{table*}[!ht]
 \caption{Neural diarization results. }
  \label{table:diarization-results2}
\vspace{-1mm}
\centering
\begin{tabular}{|l|c|c|c|c|c|c|c|c|c|}
\hline
\multirow{3}{*}{System} & \multirow{3}{*}{Data type}  & \multicolumn{7}{|c|}{DER} & \multirow{3}{*}{AVG}  \\
\cline{3-9}
& &\multicolumn{2}{|c|}{chime6} & \multicolumn{2}{|c|}{dipco} & \multicolumn{2}{|c|}{mixer6} & \multicolumn{1}{|c|}{notsofar1} & \\ \cline{3-9}
& & dev & eval & dev & eval & dev & eval & dev & \\ \hline
nsd\_{ft}\_real\_data & wpe & 12.2 & 16.2 & 14.8 & 11.3 & 7.8 & 4.4 & 8.3 & 10.7 \\ \hline
nsd\_filter\_train  & wpe & 12.4 & 15.8  & 15.7 & 10.9 & 7.9 & 5.3 & 8.9 & 10.9 \\ \hline
nsd\_{ft}\_more\_real\_data & wpe & 11.7 & 15.2 & 13.3 & 10.2 & 7.4 & 4.4 & 8.1 & 10.0 \\ \hline
nsd\_{ft}\_more\_real\_data\_er & wpe & 11.7 & 15.0 & 14.2 & 10.9 & 7.2 & 4.4 & 8.1 & 10.2 \\ \hline \hline
fusion & orig\&wpe & 10.8 & 14.8 & 13.8 & 10.0 & 7.1 & 4.3 & 7.9 & 9.8 \\ \hline
\end{tabular}
\vspace{-4mm}
\end{table*}


ND models were trained with both the original code and our re-implementation of it, keeping the original architecture but with minor changes of dataloader and optimizer. Models also differ from each other in the number of epochs for pretraining and finetuning as well as a set of channels selected for training. 
NSD-MS2S models uses the input segmentation to compute activity masks and internal speaker representations. To train such a model one needs a segmentation along with speaker labels. The best pretrained NSD-MS2S model was trained on a segmentation obtained as a result of applying clustering-based diarization from our CHiME7 pipeline \cite{CHiME7ours} with \textbf{oracle} speakers count to the simulated dataset.
For the finetuning, several segment boundaries augmentation approaches were tested. One of them uses segments from clustering-based diarization whose speaker labels are mapped to the reference ones according to the best permutation. Another one just truncates each segment by a predefined margin (0.5s) from the both sides (we call this ``bounds erosion''), surprisingly, this provides one of the best finetuning results.

\subsection{Fusion and postprocessing of the diarization results}
Diarization results are fused using the DOVER-Lap \cite{doverlap} tool in several stages. First, different clustering-based diarization pipelines are applied to each selected audio channel separately. They use 2 different Voice Activity Detectors, 2 versions of recordings (original and WPE-dereverberated) and 3 different values for the small cluster rejection parameter, 12 pipelines in total. Clustering-based diarization results are fused over all 12 pipelines in each channel separately. 
The diarization results obtained with the best single clustering-based diarization pipeline and fusion of all pipelines are presented in Table~\ref{table:diarization-results}.

Obtained clustering-based diarization results for each channel are fed into the several different neural diarization models along with either original or dereverberated audio and all results are fused again for each channel separately. Finally the diarization results from different channels are fused together.

Fusion described above provides hard labels of speaker activity. 
But speaker separation may benefit from using soft activities predicted by ND models. Therefore, a soft version of the fusion was implemented as well. For each channel we keep only those ND models whose speaker count predictions match the clustering-based diarization results. Then, the best speaker permutation between each ND model and DOVER-Lap results is found based on activities correlation. Finally all speaker-permuted activities from each selected model are averaged.

Seven ND models were trained and applied for both original and WPE-processed recordings, resulting in 14 different sets of neural diarization results. Seven results corresponding to four different NSD systems were selected for fusion based on DER minimization on the dev sets. DER values for these wpe-based variants and their fusion are presented in Table~\ref{table:diarization-results2}. The first model was trained in the original pipeline and finetuned for 6 epochs on incomplete CHiME8 data (before releasing of 2nd and 3rd parts of NOTSOFAR-1). Other 3 models were trained in our version of NSD-MS2S code. The second model was finetuned on the results of clustering-based diarization as a reference. Since the clustering-based diarization was imperfect, the finetuning was made  on a subset of CHiME8 data filtered with 2 criteria: session-channel pairs with DER $>50$\% or those where the number of speakers in clustering-based diarization results was less than the oracle one, were rejected. The third and fourth models were finetuned on the complete CHiME8 training data. 
The only difference between them was using bounds erosion in the fourth model. 

\section{Speech Separation}
\label{sec-SS}
\subsection{Guided source separation}
\label{subsec-GSS}
Our system basically uses Guided Source Separation (GSS) \cite{GSS} for the extraction of each speaker's speech based on hard or soft speech activities obtained from Neural Diarization stage.
The GSS module is based on gpu-GSS \cite{GSS-GPU}  and is used with mostly the same parameters as in CHiME7 \cite{CHiME7ours}. It includes multichannel WPE dereverberation, training complex Angular-Central GMMs (cACGMMs) for each frequency bands and using obtained source masks for MVDR beamforming. 

In the current challenge we used only the version of GSS based on the soft speaker activities. Each segment for GSS processing was taken from the results of the neural diarization and extended by a small margin (0.5s) to both sides. GSS on these extended segments was initialized by soft activities from the ND results. And the best results were obtained when GSS was run for only 5 iterations. 

There are sessions where speakers move a lot. To handle these situations a special chunked version of GSS was also used where long segments are processed by chunks of length 300 frames (about 5s). Since each of GSS versions is better on some sessions and worse on others, both of them were used in the subsequent ASR to complement each other in fusion. 

One more GSS version was used as well. Speech segments obtained from GSS described above, were recognized 
and ASR results were used to infer a (0/1) VAD masks. This mask was multiplied by soft activities to suppress noises to which ND had assigned a high activity scores. We use the same mask correction technique for the TSE model described in Section \ref{subsec-TSE}.

\subsection{Continuous source separation}
\label{subsec-CSS}
We also made an investigation of using multichannel Continuous Speech Separation (CSS) for speaker separation. We chose 2 different architectures of multichannel CSS models, namely Spatial Net \cite{SpatialNet} and TF-GridNet \cite{TFGridNet}. Since there is no clean single-speaker speech is available for CHiME8 data, we had to simulate similar training data. Clean speech recordings were selected from both LibriSpeech and VoxCeleb1,2 datasets. They were reverberated with multichannel RIRs described in subsection \ref{subsec-ND}. All training examples had a length of 4s and included 0 to 3 speakers but with no more than 2 speakers overlap, to limit CSS model by 2 output channel.
10 channel mixtures were generated and 6 best channels were selected according to MicRank to use in training. 
Targets were the clean single speaker parts of mixture convolved with respective RIRs. 

Although both models worked well on synthetic data and relatively simple mixed speech chunks, there were multiple channel confusion errors in more complex situations. This makes such an approach impractical for the application in CHiME8 challenge, especially in the scenario where CSS precedes diarization. Therefore we refused to make complete CSS. Nevertheless, application of CSS to the single-speaker segments of the CHiME8 training data resulted in a set of denoised data which was added to ASR model training and provided some improvements of the ASR results.

\subsection{Target speaker extraction}
\label{subsec-TSE}
We decided to replace CSS with Target Speaker Extraction (TSE). This task is simpler and is more relevant to the task, especially in conversations with highly overlapped speech. Besides, it doesn't require a permutation resolution and thus a model can be trained without PIT. We kept the architecture of models mostly unchanged, just extended it a bit to process a target speaker information along with mixture itself. We tried to use several representations of target speaker information, for example target speaker embeddings and target speaker activity time masks as proposed in \cite{TSEmask}. Unfortunately, ASR results on TSE outputs were inferior to those obtained on GSS+beamforming. The best ASR results on TSE outputs were obtained when the time-frequency masks from GSS were used as a target speaker representations. But they were still behind GSS+beamforming pipeline despite the good quality of target speaker's speech in most of the TSE outputs and low SNRs. 

We assumed that this gap is due to a mismatch between losses used in TSE (si-sdr) and ASR (CTC+attention). That's why we concatenated the pretrained TSE and ASR models and finetuned TSE one on ASR criterion. The TSE model was based on hard activity masks available from the ND. 
The ASR model was based on Wavlm \cite{wavlm}, see section~\ref{sec-ASR}. Since we didn't need TSE targets in this approach but only reference transcripts, this combined TSE-ASR model may be trained on CHiME8 data directly. This is an important merit of such approach. 

\begin{table}[!h]
\vspace{-2mm}
\caption{ASR results on CHiME8 devsets}
\vspace{-2mm}
\label{table:asr_results_dev}
\centering
\begin{tabular}{|ccccc|}
\hline
\multicolumn{5}{|c|}{tcpWER, \%}                                                                                                                         \\ \hline
\multicolumn{1}{|c|}{chime6} & \multicolumn{1}{c|}{dipco} & \multicolumn{1}{c|}{mixer6} & \multicolumn{1}{c|}{notsofar1} & \multicolumn{1}{l|}{MacroAvg} \\ \hline
\multicolumn{5}{|c|}{Constrained LM track}                                                                                                               \\ \hline
\multicolumn{1}{|c|}{22.79}  & \multicolumn{1}{c|}{28.96} & \multicolumn{1}{c|}{10.14}  & \multicolumn{1}{c|}{19.07}     & 20.24                         \\ \hline
\multicolumn{5}{|c|}{Unconstrained LM track}                                                                                                             \\ \hline
\multicolumn{1}{|c|}{22.47}  & \multicolumn{1}{c|}{28.41} & \multicolumn{1}{c|}{9.85}  & \multicolumn{1}{c|}{18.72}     & 19.86                         \\ \hline
\end{tabular}
\vspace{-2mm}
\end{table}

To prevent TSE from learning to reproduce the ASR training dataset, a less overfitted ASR model was chosen. 
The ASR model was trained using frozen wavlm on the reverberated Librispeech and CHiME8 training data.
We noticed that GSS is very effective at targeting the speaker, because it uses a large context. In contrast, TSE works with smaller chunks and often misses the speaker. 
For the fine-tuning, the chunk size of the TSE model was increased to a maximum of 8 seconds. Utterances longer than 8 seconds were split into chunks of 6 seconds without overlapping. However, it is still much less than the GSS context.
To improve accuracy of the TSE model, a signal from the GSS was used as a reference channel in the TSE input. As a result, TSE is in fact trained to be a GSS signal enhancer for downstream the ASR task, so we dubbed this model the Guided Target Speaker Extractor (G-TSE).

This additional finetuning significantly improved the performance of the TSE model on real-world data. The G-TSE model is a part of the best pipeline and outperforms GSS in different scenarios. We used outputs of G-TSE part as just an alternative to GSS and pass them to different ASR backends, described below.

\section{ASR and post-processing}
\label{sec-ASR}
\subsection{Speech recognition}
The speech recognition module is similar to that used in our CHiME7 system. It includes several models, both DNN-HMM hybrids trained using Kaldi \cite{Kaldi}, and a bunch of E2E models trained with either ESPNet \cite{espnet} or k2\footnote{https://github.com/k2-fsa}.
The ESPNet models have a Uconv-Conformer and an E-Branchformer \cite{E-Branchformer} architecture and use the pretrained WavLM model as a frontend. The frontend is initially frozen but then it is finetuned along with the base Uconv-Conformer model. Embeddings from the trained Uconv-Conformer models are used as features for training ZipFormer \cite{ZipFormer} model in k2.

According to our previous investigation, we trained our ASR models mainly on GSS-processed CHiME8 train datasets. But we have found that some consistent augmentation may help to slightly improve ASR accuracy. We used the clean part of LibriSpeech (460 hours) distorted with the selected RIRs (see above) and noises extracted from CHiME8 data. Another source of augmentation was the single-speaker segments of training data processed with multichannel CSS to remove environmental noise (see Section~\ref{sec-SS}).

\subsection{Adaptation}
To improve ASR results we also tried the unsupervised adaptation approach described in \cite{STAR}. The adaptation is applied on per-session level and the adapted model can provide better results compared to unadapted one. Unfortunately we didn't have enough time to try the whole adaptation pipeline but have implemented the described selection of segments for adaptation. The standard finetuning of ESPNet-based models on the selected segments brougth some improvements of the ASR results on the adaptation session. 

\subsection{Rescoring}
Each trained ASR models was applied to each variants of Speaker Separation outputs (GSS, G-TSE).
For each such combination ASR results obtained as NBest lists were re-scored with the large language model. This model was based on the Llama-2-7B\footnote{https://huggingface.co/meta-llama/Llama-2-7b-hf} LLM that was finetuned on all transcripts of CHiME8 training data augmented with LibriSpeech texts. For the finetuning, all utterances from each sessions were supplemented with BOS/EOS tokens and concatenated together in the order of ascending beginning time. During inference a context of 512 tokens composed from 1-best results of previous utterances recognition was used. This is very similar to the approach proposed in \cite{Llama-rescoring}. The rescoring results are used in the Unconstrained LM track only. 

\subsection{Fusion}
All ASR results obtained on devsets were processed to select the best subset to be fused together. All selected Nbest results were converted to Kaldi lattice format and lattice fusion along with MBR decoding were applied to obtain the final ASR results.

The tcp-WER values and macro-average over all devsets for each selected systems and the fusion results are shown in 
Table~\ref{table:asr_results_dev}.
The set of systems selected for fusion based on devsets was used to obtain the final results on eval sets. 

\section{Acknowledgements}
This work was supported by the Analytical Center for the Government of the Russian Federation (IGK 000000D730324\ P540002), agreement No. 70-2021-00141.


\end{document}